# Recycled Error Bits: Energy-Efficient Architectural Support for Higher Precision Floating Point

Ralph Nathan[1], Bryan Anthonio[2], Shih-Lien Lu[3], Helia Naeimi[3], Daniel J. Sorin[1], Xiaobai Sun[1]

[1]Duke University
[2]Cornell University
[3]Intel Labs, Intel Corporation

## Abstract

*In this work, we provide energy-efficient architectural support for floating point accuracy. Our goal is to provide accuracy that is far greater than that provided by the processor's hardware floating point unit (FPU). Specifically, for each floating point addition performed, we "recycle" that operation's error: the difference between the finite-precision result produced by the hardware and the result that would have been produced by an infinite-precision FPU. We make this error architecturally visible such that it can be used, if desired, by software. Experimental results on physical hardware show that software that exploits architecturally recycled error bits can achieve accuracy comparable to a 2B-bit FPU with performance and energy that are comparable to a B-bit FPU.*

## 1. Introduction

Society relies on the computational ability of computers for science, finance, and military applications. Software developers must consider the finite precision of floating point hardware in processor cores and the resulting potential for small inaccuracies to snowball into glaring—and silent—inaccuracies over the course of a long sequence of computations [1]. Any programmer who has ever added a very large number $X$ to a very small number $Y$ and observed the sum $X$, rather than the expected $X+Y$, is familiar with this problem.

The simplest approach to maintaining accuracy while dealing with hardware's finite precision is to use as much of that precision as possible. If the hardware supports 32-bit and 64-bit floating point, then a programmer can choose to always use the 64-bit hardware by declaring all variables as 64-bit double-precision `doubles` instead of 32-bit single-precision `floats`. In this paper, we will refer to the maximum hardware floating point precision in a processor as *M-bit precision*.

There are two problems with simply using *M*-bit precision. First, and foremost, there are situations in which even *M*-bit precision provides insufficient precision to achieve the desired accuracy. In these situations, programmers must resort to software emulation of higher precision floating point units, but this emulation incurs steep performance and energy costs.

The second problem with using *M*-bit precision is that a processor may not provide the desired performance at *M*-bit precision. Many general-purpose processors—including processors from Intel, AMD, and IBM—provide significantly greater throughput for 32-bit precision than 64-bit precision [2]. Current GPUs also tend to provide greater throughput for 32-bit precision than for 64-bit precision.[1] Programmers using such processors may prefer to use *M/2*-bit precision (32-bit in this case) for performance.

In this work, we seek to provide nearly 2*B*-bit accuracy—where *B* can be *M*, *M/2*, or some other standard precision—at a performance and energy profile that is comparable to *B*-bit hardware. We achieve this goal by enhancing the architectural interface to the core's floating point unit (FPU). Specifically, for each *B*-bit floating point addition performed, we "recycle" that operation's *error*: the difference between the finite-precision result produced by the hardware and the result that would have been produced by an infinite-precision FPU. We make this error architecturally visible as an IEEE 754-compliant floating point number that is written to a dedicated architectural register.

We call our idea "Recycled Error Bits" (REBits), and it enables a programmer to use the architecturally visible error to achieve accuracy comparable to 2*B*-bit hardware, with performance and energy that are comparable to *B*-bit hardware. REBits makes two contributions:

- REBits enables a programmer to achieve much greater accuracy than that provided by the hardware's FPU without resorting to slow and energy-intensive software emulation.
- REBits enables a programmer to use the often higher-performing *M/2*-bit floating point hardware and achieve accuracy almost as good as *M*-bit floating point hardware.

In the rest of this paper, we first explain how imprecision can lead to inaccuracy (Section 2). We then describe currently available techniques for extending precision (Section 3) and our goals for improving upon this prior work with REBits (Section 4). We then present our new architectural support for low-cost extended precision floating point math (Section 5) and how to extend floating point hardware to provide the error in each operation (Section 6). We next show how to develop software that utilizes REBits (Section 7). We then experimentally evaluate the accuracy, energy-efficiency, and performance of REBits, in comparison to existing approaches (Section 8). Lastly, we compare REBits against well-known software-only schemes to improve accuracy and show how REBits can be used to improve the performance and energy-efficiency of one such scheme (Section 9).

---

[1] http://www.nvidia.com/object/tesla-servers.html

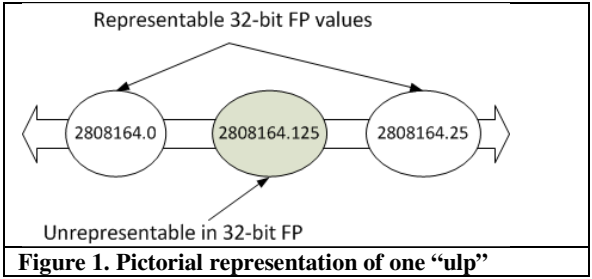

**Figure 1. Pictorial representation of one "ulp"**

## 2. Motivation

### 2.1. How Imprecision Can Cause Inaccuracy

Inaccuracy arises due to finite precision. Fundamentally, we are trying to represent an infinite number of floating point values with a finite number of bits. We assume that all floating point numbers and arithmetic adhere to the IEEE 754 standard [3], and we refer readers interested in a refresher on the floating point standard and the limitations of floating point hardware to Goldberg [4].

Computer architects and most software developers realize that the results of floating point computations are rounded. Let us assume 32-bit floating point operations for this discussion. When adding two 32-bit floating point numbers, each of which has a 24-bit mantissa (23 explicit bits plus the implicit leading "1"), the result can have a mantissa that is longer than 24 bits. When this situation occurs, the result of the mantissa is rounded to 24 bits, and thus the result is inaccurate due to the finite precision of the hardware. For example, consider the addition of 2808064.0 and 100.125. The rounded 32-bit result is 2808164.0, which reflects an error of 0.125 with respect to the true result of 2808164.125.

The absolute error of a single operation is exceedingly small: less than one <u>u</u>nit in the <u>l</u>ast <u>p</u>lace (often called an "ulp"). An ulp is the difference between the rounded result and the next nearest floating point number to it. In Figure 1, we illustrate how a given floating point number, the 2808064.125 result from our previous example, might be between two representable 32-bit floating point numbers. There is inherently some inaccuracy in rounding this value to either of its closest neighbors.

Although that inaccuracy is indeed small, inaccuracies can accumulate over the course of a long sequence of operations. One might expect these inaccuracies to cancel each other out, on average, but there are three problems with this expectation.

1. Even if a large accumulation of errors is unlikely, there are many applications for which *any* possibility is unacceptable.
2. There are many well-publicized and non-contrived examples in which errors do not cancel, causing disastrous results [1].
3. When adding two numbers of different signs but nearly identical absolute values, there exists the possibility of greatly magnifying a small inaccuracy. Consider a base-10 example in which we add $A=9.32415 \times 10^{18}$ and $B=-9.32414 \times 10^{18}$. The expected result is $1.0 \times 10^{13}$. However, if during the computation of $B$ (in previous instructions), rounding error affects the least significant bit of the mantissa of $B$ such that it rounds to $-9.32415 \times 10^{18}$, then we get a result of 0. A one-ulp inaccuracy in one operand in this example causes an inaccuracy of $1 \times 10^{13}$ in the result.[2]

We show later, using microbenchmark kernels and scientific benchmarks, that rounding error can indeed accumulate.

### 2.2. Addition, the Most Important Operation?

Software that sums a large number of values—which is a common feature in scientific code—is particularly vulnerable to accumulated rounding error. Our work specifically addresses the error that can accumulate in such sums. We consider only addition because a running sum of computed products (or quotients or numbers computed otherwise) can be considered a running of sum of numbers that are given. Another reason to consider only addition is the fact that, for rounding error to be "significant," a long computational chain needs to exist (e.g., a running sum).

## 3. Prior Work in Extending Precision

There are three well-known approaches for providing precision greater than 64-bit IEEE-754 hardware.

### 3.1. Double-Double Arithmetic

One can increase precision by using software to stitch together multiple floating point values and computations. A notable example is the double-double arithmetic developed by Bailey's group [5]. A double-double floating point value is represented by two `doubles`, and the value of the double-double is the sum of these two `doubles`. If *A* and *B* are both `double-doubles` (pairs of 64-bit values), then adding *A+B* produces a `double-double` *C*. The challenge is that adding two `double-doubles` requires more than two hardware floating point additions. In fact, adding two `double-doubles` requires 20 64-bit hardware additions. Thus `double-double` arithmetic requires substantially more work and thus takes more time and consumes more energy. Natural extensions of `double-double`, such as `quad-double`, extend the accuracy/cost trade-off. We show later, in Section 9.2, how to use REBits to accelerate Double-Double operations and make them more energy-efficient.

### 3.2. Simulation of Arbitrary-Precision FPU

An extreme way to achieve accuracy is to simulate an arbitrary-precision FPU and not use the floating point hardware at all. A well-known GNU library, called GMP (GNU Multiple Precision),[3] uses integer and bit-manipulation instructions to perform arbitrary-precision floating point computations. GMP thus provides the desired accuracy, but at an extremely large cost in terms of time and energy. We use

---

[2] To be fair, numerical analysts write numerically-aware code to try to avoid such situations, but these situations can occur, particularly for less numerically-aware programmers.

[3] www.gmplib.org

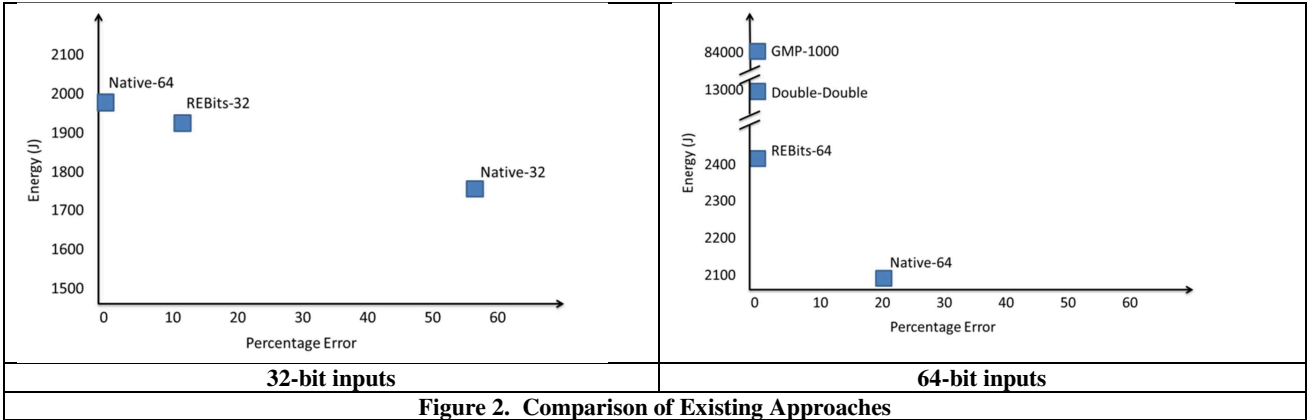

**Figure 2. Comparison of Existing Approaches**

GMP with 1000 bits of precision (which we denote as GMP-1000) as a "gold standard" of accuracy in our experiments.

### 3.3. Intel's x87 Floating Point

Intel processors support the legacy x87 floating point specification, in which floating point values within the core are 80-bit quantities and thus more precise than 64-bit `doubles`. The x87 standard does not adhere to the IEEE-754 standard, though. Also, because floating point values are 64-bit quantities in the memory system (i.e., not 80-bit), the result of a computation depends on register spills and fills and is thus not deterministic.

## 4. Goal for Recycled Error Bits

We visually represent the state of the art and our goal for REBits in Figure 2. In each plot, the x-axis is inaccuracy (percent error in the computation) and the y-axis is system energy consumption minus the energy consumed by an idle system. The goal is to get as close to the origin (bottom left) as possible. The data points "Native-32" and "Native-64" correspond to using the native FPU at 32-bit and 64-bit granularity, respectively. "Double-Double" refers to the Double-Double scheme [5] described in the previous section. "REBits-32" and "REBits-64" denote our approach (which we describe in the next section) for augmenting 32-bit and 64-bit FPUs, respectively. We only compare schemes operating on the same input sizes, because otherwise comparisons are misleading. The microbenchmark, which we describe in more detail in Section 8.2.1, is a simple kernel that performs the summation of a long sequence of random numbers. The key takeaway points—which apply to this microbenchmark as well as to several complete benchmarks—are that:

- <u>32-bit inputs (left side of figure)</u>: REBits-32 uses slightly more energy than Native-32 but achieves accuracy far closer to Native-64 than to Native-32.
- <u>64-bit inputs (right side of figure)</u>: REBits-64 uses slightly more energy than Native-64 but achieves much greater accuracy. Furthermore, REBits-64 achieves comparable accuracy to Double-Double (and, in this case, GMP-1000) while using vastly less energy; note the two discontinuities on the y-axis.

We now explain how REBits works and achieves these goals.

## 5. Recycled Error Bits

We now present our architectural support for numerical accuracy, called *Recycled Error Bits (REBits)*.

### 5.1. Big Picture

The key idea of REBits is to make the rounding error in each floating point operation architecturally visible. In this work, we consider only addition due to the prevalence of running sums in scientific, floating point intensive applications. The error is the difference between the rounded result, which is what FPUs produce today, and the result that would have been obtained with an (unimplementable) infinite-precision FPU. In today's cores, the FPU discards information when it rounds, and our goal is to recycle this information to help programmers achieve greater accuracy when desired. We illustrate this high-level view of REBits in Figure 3.

### 5.2. Architecture

REBits makes the rounding error of each floating point addition architecturally available in a dedicated register. In this work, we consider only addition; we plan to extend this work to multiplication in the future. With addition and multiplication in place, extensions to other floating point operations—such as division, square root, cosine, sine, etc.—are routines in either hardware or software.

With REBits, each floating point add instruction (fpadd) produces a sum that, as usual, is written to an architectural register specified in the instruction. The difference with REBits is that another IEEE-754-compliant floating point value, the error, is written to a dedicated architectural register we call FPERR (that is saved/restored at context switches). The value of FPERR is overwritten by every fpadd instruction. Thus, a programmer who wishes to use this information must move it from FPERR to a regular floating point register before the next fpadd instruction. If the ISA has unused register specifier bits in its floating point move instruction, then the move from FPERR is just like any other move. Otherwise, we must add a new instruction to the ISA that moves the contents of FPERR to a specified floating point register.

Many ISAs support multiple fpadd precisions, including 32-bit and 64-bit (and sometimes 16-bit). A simple approach to REBits is to have a dedicated FPERR register for each, e.g., FPERR32, FPERR64, etc.

Some ISAs also support packed floating point operations, such as Intel's SSE. (In fact, in modern x86 cores, only SSE complies with IEEE-754, and it is used exclusively by many compilers unless otherwise specified.) With packed *N*-bit arithmetic, each register is treated as *k N/k*-bit floating point values, where *k* is specified in the instruction. For example, a 128-bit register could be interpreted as 2 64-bit values or 4 32-bit values. For REBits, we can simply extend packing to the FPERR register, i.e., FPERR is interpreted as *k N/k*-bit floating point error values.

Floating point arithmetic has some edge cases such as infinity and Not-A-Number (NaN). These situations do not directly affect REBits, because the error can never be infinity or NaN. If the answer is infinity or NaN, the error does not matter. We do not currently handle sub-normal numbers ("denorms").

We envision architectural support (e.g., a mode bit) for disabling REBits when its features are not being used. Our microarchitectural changes for REBits and our FPU implementation (both discussed later) lend themselves to being power-gated to save energy when not in use.

### 5.3. Microarchitectural Design Issues

There are two microarchitectural issues that are introduced by the REBits architecture.

#### 5.3.1 Pipeline Bypassing

In a pipelined microprocessor core, the value of FPERR may need to be consumed before it has been written into its register. For example, consider the canonical 5-stage pipeline (Fetch, Decode, Execute, Memory, Writeback) of the Patterson and Hennessy textbook [6]. If a fpadd instruction is in Execute and an instruction to move FPERR is in Decode, then the pipeline must bypass FPERR from the fpadd to the move. This type of bypassing is straightforward, and we mention it for completeness rather than because it is an obstacle.

#### 5.3.2 Register Renaming

The other microarchitectural issue introduced by REBits is incorporating REBits within an out-of-order microprocessor core. Architecturally, REBits specifies that the current value of FPERR is the error of the most recent fpadd *in program order*. However, an out-of-order core does not necessarily execute instructions in program order.

Our relatively straightforward solution to this problem is to extend register renaming to include FPERR. The only subtle difference—which is also an issue for condition flag renaming [7]—is that the fpadd instruction that produces FPERR as its result does not *explicitly* specify FPERR as an output. We extend the renaming logic to rename FPERR on each fpadd instruction.[4] Thus, when a move instruction accesses FPERR as an input operand, it accesses the renaming logic to obtain the most recent value (in program order) of FPERR. To support renaming of FPERR in pipelines with significant pressure for floating point registers, it may be necessary to provide additional physical floating point registers; the number of extra physical floating point registers desired would depend on the expected number of fpadds in flight.

---
[4] A more efficient approach would allow the compiler to mark only those fpadd instructions whose FPERR will be read and then the hardware would rename only the marked fpadd instructions.

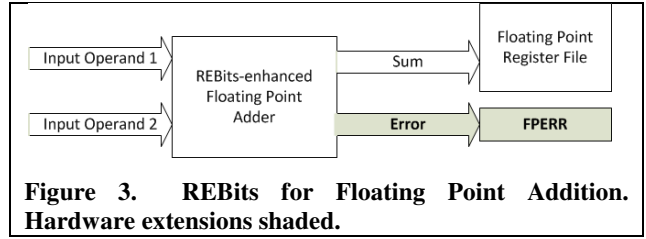

**Figure 3.** REBits for Floating Point Addition. Hardware extensions shaded.

<u>Disclaimer</u>: We do not quantify the microarchitectural impact of REBits since that would require implementing this design in a superscalar processor.

## 6. FPU Implementation

The key hardware innovation of REBits is to extend the functionality of the floating point adder such that it recycles the error that is normally discarded. Recycling is an apt analogy, because most of the logic required to determine the error is already in the FPU and we simply need to maintain some bits that are otherwise ignored.

### 6.1. Where Error Occurs

During the course of a floating point addition, there are two steps in which error is introduced and must be tracked by our extended FPU. Assume that we are adding two numbers *A* and *B* and, for ease of exposition, both numbers are positive and *A* has a greater exponent than *B*. We illustrate a 16-bit example in Figure 4, and we shade the two steps where error is introduced.

The first step in which error is introduced is when *B* is denormalized such that its exponent is aligned with *A*'s exponent. This step is the second column from the left in the figure. During this step, *B*'s mantissa is shifted to the right, and any number of *B*'s mantissa bits may be lost.

The second step in which error can be introduced is during the final rounding step (rightmost column in the figure). During the course of the addition operation, the intermediate, pre-rounded result can use up to 28 bits of precision in the mantissa, yet the final result must use only 24 bits of mantissa (23 plus the implicit "1"). To trim the mantissa down to 24 bits, the FPU rounds the result according to the IEEE standard. The difference between the pre-rounded and post-rounded result is additional error.

### 6.2. Recycling Error

To recycle error, our FPU must determine the error introduced in the two steps of the addition described in the previous section. We illustrate how the 32-bit hardware works in a simplified flow chart in Figure 5. This figure walks through the process of adding two `floats`, *A* and *B*, assuming both are positive and *A*>*B*. Steps 2 and 4 in the figure correspond to the two steps in which error is introduced (i.e., the shaded entries in Figure 4).

| A=1.1101001101x2^14<br>B=1.1000111011x2^10 | A=1.1101001101000x2^14<br>B=0.0001100011101x2^14 | sum_pre=1.1110110000101x2^14 | sum=1.1110110001x2^14 |
|---|---|---|---|
| *original operands to be added* | *denormalized B, lost least significant bit of B* | *pre-rounded result has 15 bits of precision in mantissa* | *rounded result, lost some accuracy* |

**Figure 4. Where error is introduced during floating point addition**

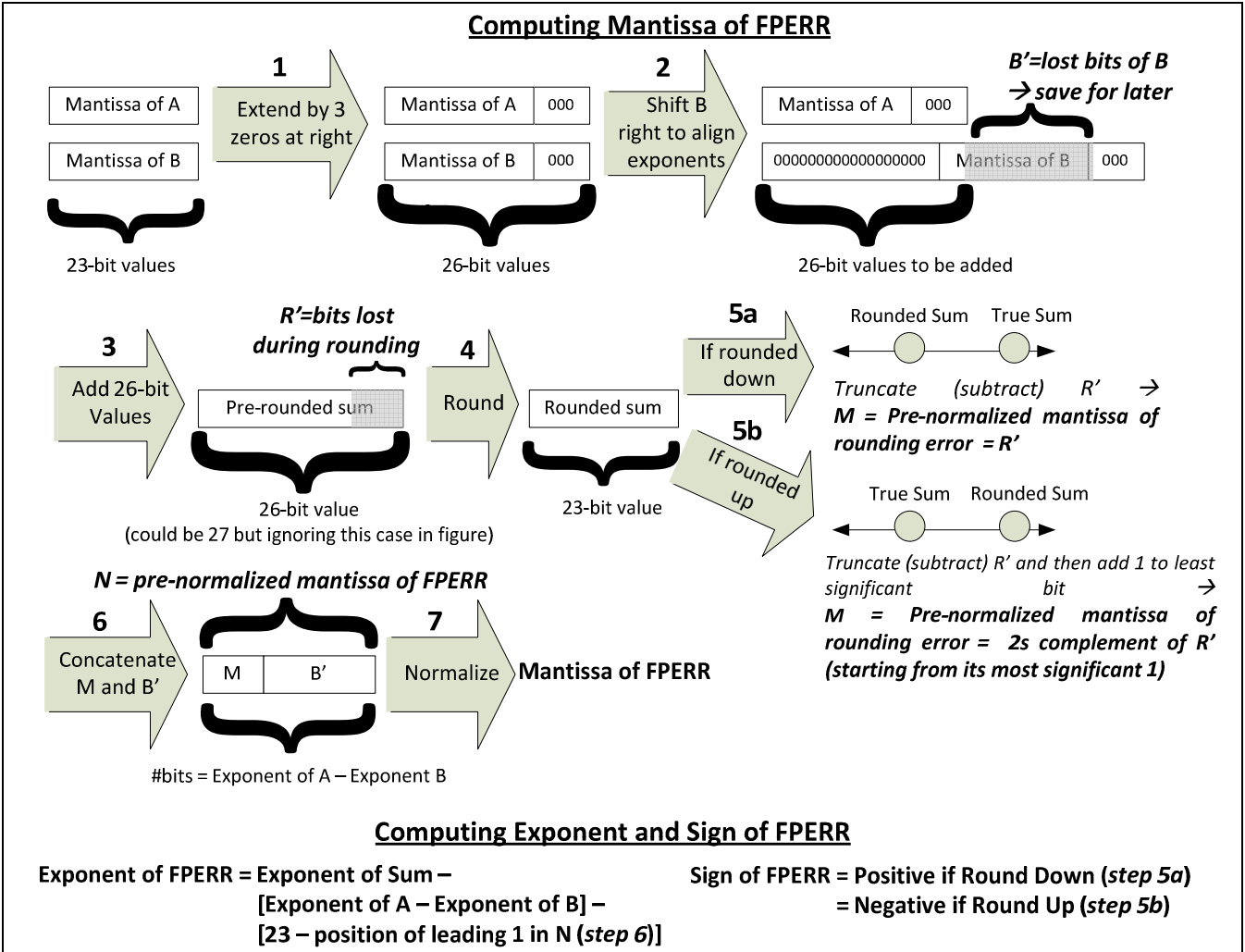

**Figure 5. Hardware for Computing FPERR. Assume A and B are positive and A>B.**

| ```
float result = 0;
float[N] v;
for (i=0; i<N; i++){

result=result+v[i]*v[i];
}
result = sqrt(result);
``` | ```
float result = 0;
float err = 0;
float[N] v;
for i=0; i<N; i++){

result=result+v[i]*v[i];
   err=err + FPERR;
}
result = result + err;
result = sqrt(result);
``` | ```
floaterr result;
result.val = 0;
result.err = 0;
float[N] v;
for i=0; i<N; i++){
   result = result + v[i]*v[i];
}
result.val=result.val+result.err;
result = sqrt(result);
``` |
|---|---|---|
| **Naïve accuracy-unaware code** | **REBits with global variable "FPERR"** | **REBits with new datatype "floaterr"** |

**Figure 6. 2-Norm Software. REBits extensions highlighted in bold text.**



The process of obtaining the error may appear somewhat complicated, but the actual hardware required for this purpose is relatively simple. The hardware includes some registers, shifters, bit manipulation logic, and combinational logic for control.

### 6.3. Hardware Implementation

We implemented the REBits-32 floating point adder as an extension of an open-source 32-bit floating point adder from OpenCores.[5] We implemented the adders in synthesizable Verilog using Synopsys CAD tools and 45nm CMOS technology from Nangate [8]. To accurately determine energy, we back-annotated the netlists with the parasitic resistances and capacitances, and we simulated the circuits' operation on actual floating point inputs.

The three potentially relevant issues for the hardware implementation are its latency, area, and energy consumption. As we show later (Section 8.2.1), the energy consumption of the floating point adder itself is negligible; the system-wide energy difference between using the 32-bit and the 64-bit adder is far less than 1%. (The system-wide energy difference between 32-bit and 64-bit floating point derives from the energy in transferring data to/from memory, not from the computations themselves.) Thus, the energy consumption of the REBits-32 floating point adder is negligible as long as it is less than the baseline 64-bit adder.

In Table 1, we present the overheads—latency, area, and energy—for REBits-32. In all three categories, REBits-32 is between the baseline 32-bit and 64-bit adders, yet significantly closer to the 32-bit baseline. REBits-32 consumes less energy than the 64-bit baseline, so it indeed has negligible impact on system-wide energy. REBits-32 has a roughly 10% latency overhead that could be further reduced with more aggressive pipelining. REBits-32 has an area overhead of about 2500 μm$^2$, which is a fairly small fraction of the entire FPU.

**Table 1. Hardware Overheads: Latency, Area, Energy**

|  | Baseline 32-bit | REBits 32-bit | Baseline 64-bit |
|---|---|---|---|
| Latency | 4.45 ns | 4.89 ns | 8.03 ns |
| Average energy | 7.72 pJ | 8.40 pJ | 13.9 pJ |
| Area - FP adder | 4804 μm$^2$ | 7384 μm$^2$ | 9560 μm$^2$ |
| Area – entire FPU | 15902 μm$^2$ | 18165 μm$^2$ |  |

## 7. Utilizing REBits in Software

The goal of this work is to provide low-cost architectural support for extending the precision of floating point. So far we have described the architectural support, and now we turn our attention to how software uses it.

### 7.1. Programming Interface

There are several ways in which we could make FPERR visible to an application programmer.

- *Inline assembly code* may not be the most programmer-friendly option, but most programmers of numerical software are sophisticated enough to use it.
- *Compiler intrinsics* are perhaps more programmer-friendly than assembly, but they are still not going to be used by inexperienced programmers.
- A *language-level global variable* that holds FPERR would be easy to use, and it is easily ignore-able by programmers uninterested in numerical issues.
- A *language-level datatype* that is a struct that contains both the value and the error would also be easy to use. For example, we could create a `floaterr` datatype with floaterr.val and floaterr.err. Consider the addition of two `floaterrs`, v1 and v2, to produce a `floaterr` sum v3, i.e., v3 = v1 + v2. The semantics of this operation are that v3.val = v1.val + v2.val and v3.err = v3.err + FPERR.[6]

We do not strongly advocate for one option among these choices, but we believe that the last two options are probably the most likely to be adopted. In the next section, we present software that uses both options.

### 7.2. Example REBits Software

We now illustrate how we use REBits in actual software, using a simple example: a function that computes the 2-norm of a vector of numbers:

$$2norm(\vec{V}) = \sqrt{\sum_i V_i^2}$$

In Figure 6, we show three implementations of 2-norm code. From left to right, we present the accuracy-unaware code, REBits code with a global FPERR value, and REBits code with a `floaterr` datatype. In the REBits code, we highlight REBits extensions with bold text.

Both versions of the REBits code require minimal modifications to the accuracy-unaware code, and understanding these modifications does not require an advanced degree in numerical analysis.

### 7.3. Inappropriate Software for REBits

We do not claim that REBits is easily applicable to all numerical software. We now highlight three classes of algorithms for which REBits is either unhelpful, insufficient, or memory-intensive.

<u>Unhelpful:</u> REBits is most helpful when the algorithm performs a summation of a large number of floating point values. Such summations are common in scientific simulations [9] and other floating point software, but they are not ubiquitous. Code without summations is unlikely to benefit much from REBits.

<u>Insufficient:</u> Another limitation of REBits is that it cannot overcome gross accuracy problems; it was not designed for this purpose. If software performs frequent floating point division (or other native hardware operations with known inaccuracy, such as square root, sine, etc.), then the

---

[5] http://opencores.org/project,fpu100

[6] Another reasonable semantic would be: v3.err = FPERR.



inaccuracies introduced during divisions are likely to far outweigh any accuracy benefits derived from REBits. Anecdotally, we optimistically attempted to apply REBits to a Taylor series expansion of sin(x), but the division in each term in the series created inaccuracies that dwarfed any possible inaccuracies incurred when adding the terms together.

Memory-Intensive: There are algorithms for which REBits is useful but requires significantly more memory than non-REBits software. In general, there exist applications which require more than one running sum of error . (Compare to the 2-norm algorithm in Figure 6, which requires only one variable to hold the running sum of error.) These algorithms require vectors or arrays of errors, and these vectors or arrays can be as large as the actual working sets. Such algorithms include FFT, 1D diffusion simulation, and Strassen's algorithm for matrix multiplication [10].

## 8. Experimental Evaluation

We now experimentally evaluate REBits, both on kernels and then on complete benchmarks. The kernels are chosen to highlight and isolate the specific aspects of numerical codes that we focus on in this work, especially the summation of a long sequence of numbers. The results on the easily analyzed kernels provide insights. The kernels are not contrived—they represent snippets of actual numerical software—but they are not complete programs. The complete benchmarks are chosen to be representative of typical, commonly-used numerical software, and the results on these benchmarks are what ultimately matter.

### 8.1. Methodology

We evaluate REBits in comparison to existing hardware and software (denoted "Native"), and we compare 32-bit and 64-bit versions: Native-32, Native-64, REBits-32, and REBits-64. We evaluate accuracy, performance, and energy consumption. All of our kernels and benchmarks, except one, have 32-bit inputs; for these experiments we do not evaluate the accuracy of REBits-64, which is overkill. For the kernel with 64-bit inputs (Section 8.2.3), we compare Native-64 to REBits-64.

We run all experiments on actual (not simulated) x86 hardware, using only the SSE units (not x87). The system is an Intel Core i5-2500 with 4 cores, a 3.30 GHz clock, 8 GB of DDR3 DRAM, a Radeon HD 5450 GPU, and a 250 GB Seagate hard drive. The system has an idle energy consumption of 48.5W, as measured at the wall outlet. The system runs Linux 3.0.0-22-generic. All software is compiled with gcc 4.6.1 and –O3 optimization level.

Accuracy: We compare the numerical results for existing hardware/software and REBits to numerical results produced by the arbitrary-precision GMP library using 1000 bits of precision. GMP-1000 is the "gold standard" of accuracy. To simulate the REBits FPU, we use the open-source SoftFloat[7] software simulator of an IEEE-754 compliant FPU. The FPU's adder supports all rounding modes supported by IEEE-754. We modified the SoftFloat FPU's adder to produce FPERR.

Performance: We measure runtimes using hardware performance counters. For evaluating the runtime of REBits code, we must consider the latency required for moving FPERR to another register, yet our actual hardware does not have an FPERR register. We mimic the latency of the instruction required to perform this move by inserting an inline assembly instruction to move one floating point register to another floating point register, such that the functionality of the program is not affected and while preserving the dataflow as if we had the actual error register. This approximation of the performance of REBits is disadvantageous to REBits, because the compiler is constrained in optimizing around the inline assembly.

Energy: We measure power with a meter at the wall outlet. Power thus includes all aspects of the system, including the disk drive, power supply, fan, etc. We sample the power readings from the meter every second with an accuracy of 1.5%. We subtract the idle power of our system, 48.5W, from all power readings to obtain what we refer to as the "active power" of our experiments. We use the trapezoid method to integrate the active power samples to compute the active energy consumed. (We call "active energy" simply "energy" in the rest of this paper.)

We report total system energy—and not which part of the hardware consumes the energy—because most of the differences in energy consumption are due to memory accesses. We performed identical experiments on 32-bit and 64-bit data and, if the dataset fit within the cache, the experiments consumed essentially the same energy (<1% difference). These experiments demonstrate that the system's energy consumption does not vary as a function of how the core internally performs floating point computations. (Energy results differ greatly, however, when the dataset does *not* fit in the cache.) Furthermore, because the energy consumed by the FPU itself is a negligible fraction of the total energy, we do not model the extra energy consumed by the REBits FPU during each fpadd.

As with the performance experiments, we mimic the instruction to move FPERR to another register using inline assembly, and thus we closely approximate the energy consumption of the move that would occur in REBits. We run each experiment 10 times to account for the possibility of variability across experiments, although, in practice, we observe much less than 1% variation across experiments.

### 8.2. Kernels

We analyzed three kernels to isolate exactly those software idioms that are improved by REBits. These kernels are not contrived examples; rather, these kernels are commonly used in complete benchmarks.

---

[7] http://www.jhauser.us/arithmetic/SoftFloat.html



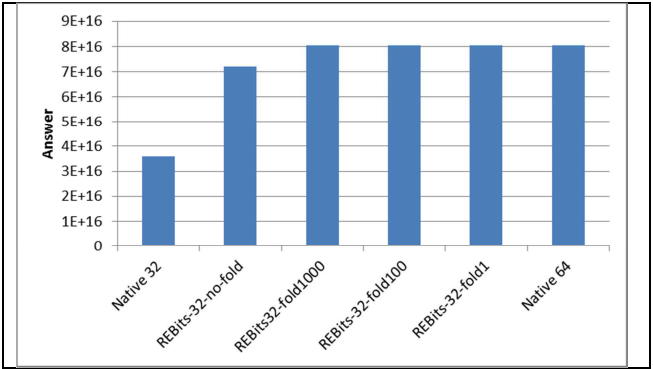

**Figure 8. Summation of Positive Numbers: Accuracy**

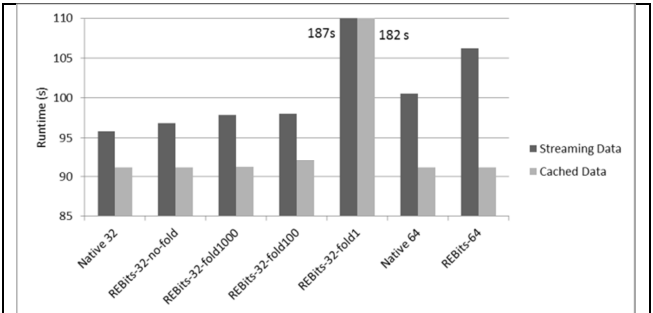

**Figure 9. Summation of Positive Numbers: Runtime**

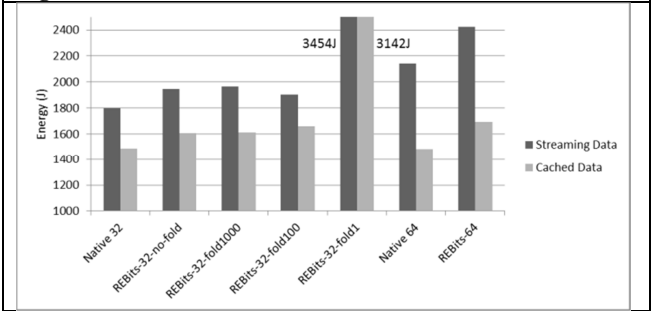

**Figure 10. Summation of Positive Numbers: Energy**

### 8.2.1 Summation of Sequence of Positive Numbers

We perform the summation of $N$ random positive floating point numbers, and we choose the $N$ values such that the first ¾ are very large and the last ¼ are very small. Each experiment consists of a large number, $R$, of computations of the sum, to minimize the effects of initialization. To enable fair comparisons for the two values of $N$, we choose $R$ such that $N \times R$ is a constant (one billion). We perform each experiment multiple times with different random seeds, and we observe negligible differences between experiments.

We run with two different values of $N$. We use $N$=100,000 to isolate effects on the core, because this sequence size fits comfortably in the core's L3 cache (i.e., does not access memory). We use $N$=100,000,000 to examine the effects on memory, because this size exceeds the L3 and streams data from memory.

For REBits, we use the algorithm in Figure 7. We parameterize the algorithm by how often we "fold in" the running error. As the program runs, the running error can become large, and accuracy is improved if we take the part of the running error that is as significant as the running sum and add them together. The part of the running error that is less significant remains in the running error. As we fold in the error more frequently, we improve accuracy but at the cost of increased latency and energy.

We show the accuracy of three schemes—Native-32, Native-64, and REBits-32—on this kernel (with the larger value of $N$=100,000,000) in Figure 8. Native-64 is equal to GMP-1000, as well as to REBits-64, so we do not show either of those. The REBits-32 datapoints are labeled with the frequency of error folding (FoldErr in Figure 7). The REBits-32-no-fold datapoint denotes no folding within the main loop and just a single folding in of the error after the entire loop completes. The results show that all versions of REBits-32 are much more accurate than Native-32, and that folding even once per 1000 iterations obtains the same result as Native-64.

The question now is what cost we pay to achieve these accuracy gains over Native-32. In Figure 9, we present runtime results for both values of $N$ (denoted "Streaming Data" and "Cached Data"). The runtime overheads for REBits are due to the instructions to move FPERR to a register and folding, and these overheads are generally quite small. In particular, Native-64 takes noticeably longer to run than REBits-32-fold1000, despite having the same accuracy.

In Figure 10, we present the energy results. As with runtime, the differences for the "Cached Data" are fairly small, except for when we fold every iteration. In fact, Native-32 and Native-64 are almost identical, which implies that the energy consumption of the floating point adder itself has negligible impact on the system's total energy. For the "Streaming Data," there are some significant differences. Most importantly, Native-64 consumes much more energy than Native-32 and REBits-32 for modest amounts of folding. Thus, REBits-32 achieves comparable accuracy to Native-64 while using far less energy. One curious and counter-intuitive result is that REBits-32-fold100 consumes slightly *less* energy than REBits-32 when it folds less often. We do not yet have a good explanation for this data.

### 8.2.2 Parallel Summation of Positive Numbers

To study the impact of multithreaded software, we parallelized the kernel from the previous section using pthreads. The results, provided in Table 2, are more dramatic.

```
float sum=0, err=0;
float[N] v;
const int FoldErr; // how often to fold
for i=0; i<N; i++){
   sum = sum + v[i];
   err = err + FPERR;
   if (i % FoldErr == 0){
       sum=sum+err;   // fold in error
       err = FPERR;
   }
}
result = result + err;
```

**Figure 7. REBits Code for Sum of Positive Numbers**



**Table 2. Parallel Summation Results**

|  | Sum | Runtime (s) | Energy (J) |
|---|---|---|---|
| Native-32 | 897431984 | 27.26 | 1412 |
| REBits-32 | 976661328 | 27.42 | 1481 |
| Native-64 | 968499228 | 44.39 | 2702 |
| GMP-1000 | 968499228 |  |  |

REBits-32 (folding once at the end of the loop) has an error of less than 0.84%, whereas Native-32 has an error of 7.3%. The runtime and energy of REBits-32 are nearly identical to Native-32, yet Native-64 has runtime and energy overheads of 63% and 91%, respectively. These larger overheads for the parallel summation reflect the memory pressure exerted by the multiple threads. Native-64 transfers twice as much data to and from memory, and these transfers are a performance bottleneck and energy drain. Most of the core time is spent waiting for memory to provide data. The reason that the speed-up when going from serial to parallel for Native-64 is less than that of Native-32 is due to the fact that our system is bandwidth limited in that case.

*8.2.3 Summation of Sea Heights*

A prior paper [9] on numerical accuracy in scientific simulations, focusing on the problem of summations, presented a stress-test kernel that sums a remarkably small, yet problematic, 2D (120x64) array of sea heights. This kernel is similar to the one in the previous section, except the data is real (not randomly generated) and has both positive and negative values. The authors observed that the answer differed greatly depending on the order in which the sea heights were summed (e.g., row-first vs. column-first). In our experiment, we perform the summation one million times to minimize the effects of initialization (i.e., declaring the array of sea heights, reading the array entries in from a file, etc.). This is the only kernel or benchmark in this paper with 64-bit inputs, so we compare Native-64 and REBits-64.

In **Table 3**, we present the experimental results. The "Sea Height" column illustrates the accuracy issues for this kernel. The Native-64 results indeed vary wildly based on the order of summation, and none of the summation orders lead to correct results. Some Native-64 results are incorrect by two orders of magnitude. REBits-64, however, obtains the same result in all orders of summation, and this result is the same as GMP-1000. The runtimes and energy consumption of REBits-64 are comparable to or better than Native-64, depending on the order of summation. Thus, for comparable costs, REBits achieves far better accuracy than native hardware/software. At the other end of the cost spectrum, Double-Double and GMP-1000 have runtimes and energy consumptions that are orders of magnitude greater than Native-64.

*8.2.4 2-Norm*

The 2-norm kernel was previously described in Section 7.2. We perform the experiment in the same fashion as the experiment in Section 8.2.1, with the same vector length and the same method for choosing the vector elements. We also perform multiple repetitions of the computation to minimize initialization effects.

**Table 3. Sea Heights: Experimental Results**

|  | Sea Height | Runtime (s) | Energy (J) |
|---|---|---|---|
| Native-64 Row-first | 34.418 | 6.99 | 314.6 |
| Native-64 Reverse-row-first | 32.3027 | 6.99 | 315.6 |
| Native-64 Col-first | 0.48759 | 7.025 | 406.7 |
| Native-64 Reverse-col-first | 0.16016 | 7.404 | 402.3 |
| REBits-64 | 0.35799 all directions | 6.99 row-first | 321.0 row-first |
| Double-Double | 0.35799 all directions | 56.6 row-first | 833.1 row-first |
| GMP-1000 | 0.35799 all directions | 200.2 row-first | 4120 row-first |

The experimental results are in Table 4. The actual 2-norm result produced by GMP-1000 and by Native-64 is $1.074 \times 10^{13}$. REBits-32 produces a result of $1.068 \times 10^{13}$, which represents a small percent error of 0.6%. Native-32, however, produces a result of $8.796 \times 10^{12}$, which is a percent error of 18%.

The runtime overheads of REBits-32 and Native-64, compared to Native-32, are <1% and 7%, respectively. The energy overheads are 3.4% and 18%, respectively. Once again, REBits-32 is producing results comparable to Native-64 at costs that are comparable to Native-32 and far less than the costs of Native-64.

**Table 4. 2-Norm: Experimental Results**

|  | 2-Norm Value | Runtime (s) | Energy (J) |
|---|---|---|---|
| Native-32 | $0.8796 \times 10^{13}$ | 97.9 | 2056 |
| REBits-32 | $1.068 \times 10^{13}$ | 97.9 | 2125 |
| Native-64 | $1.074 \times 10^{13}$ | 104.9 | 2429 |
| GMP-1000 | $1.074 \times 10^{13}$ |  |  |

**8.3. Benchmarks**

We identified several important, commonly-used scientific benchmarks that perform extensive summations.

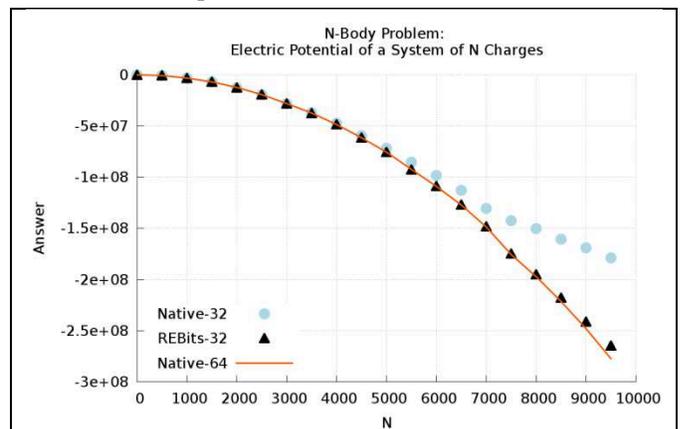

**Figure 11. N-Body Simulation: Accuracy**



### 8.3.1 N-Body Simulation of Electrical Charges

Many scientific applications are examples of N-body simulations, where bodies can be molecules, planets, etc. An example well-known to computer architects is barnes-hut, a benchmark in the Splash-2 benchmark suite [11]. This particular N-body simulation calculates the electric potential of a system of point charges. The position and the magnitude of the charge of each particle are randomly determined at the beginning of each simulation. The benchmark is parameterized by *N*, the number of charges in the system. We started with code that Prof. Richard Gonsalves uses in his class on computational physics at the University of Buffalo and that he generously provided to us.

In Figure 11, we show the accuracy of Native-32, Native-64, and REBits-32, all as a function of *N*. Native-64 is indistinguishable from GMP-1000, so we do not show GMP-1000. We observe that Native-32 and REBits-32 provide excellent accuracy for *N*<5000. At approximately *N*=5000, Native-32 quickly loses accuracy. REBits-32 maintains accuracy throughout, with a maximum percentage error of 4.5%, which is far less than the maximum of 35.5% for Native-32.

The performance and energy of the three schemes are in Table 5. The results show that REBits-32 has a runtime and energy consumption that are quite close to Native-32. Native-64 has runtimes and energy consumptions that are much greater. In particular, Native-64 uses 45% more energy than Native-32, whereas REBits-32 uses less than 1% more energy.

**Table 5. N-Body Simulation: Runtime and Energy**

|          | Runtime (s) | Energy (J) |
|----------|-------------|------------|
| Native-32 | 89.6       | 1570       |
| REBits-32 | 90.8       | 1584       |
| Native-64 | 119.7      | 2280       |

### 8.3.2 Numerical Integration

Numerical integration is used extensively in scientific computing. We used the Trapezoid Rule algorithm to numerically integrate the following equation:

$$y = \int_0^x 400(x \sin(x) + \cos(x) - 1)$$

In Figure 12, we show the accuracy of Native-32, Native-64, and REBits-32. Native-64 is nearly identical to the golden standard of GMP-1000, so we do not plot GMP-1000. For clarity, we plot every value of *x* for Native-64, but we plot only a periodic sample of *x* values for Native-32 and REBits-32. (If we plot all values of *x* for Native-32 and REBits-32, the graph is illegible.) We observe that REBits-32 has accuracy that is very close to that of Native-64. The largest percentage error for REBits-32 is 4.0%, which is many orders of magnitude smaller than those of Native-32, which are as large as 60,000%.

Table 6 shows the runtime and energy results. We observe that the runtime and energy for REBits-32, Native-32, and Native-64 are all quite similar. Thus, on a processor without a 64-bit FPU, REBits-32 is an attractive option.

**Table 6. Numerical Integration: Runtime and Energy**

|          | Runtime (s) | Energy (J) |
|----------|-------------|------------|
| Native-32 | 85.4       | 1434       |
| REBits-32 | 86.0       | 1459       |
| Native-64 | 86.8       | 1460       |

**Table 7. Financial Simulation: Runtime and Energy**

|          | Runtime (s) | Energy (J) |
|----------|-------------|------------|
| Native-32 | 159.2      | 2884       |
| REBits-32 | 157.2      | 2847       |
| Native-64 | 161        | 2815       |

### 8.3.3 Financial Pricing Simulation

Many financial applications rely on floating point arithmetic to predict prices for stocks, bonds, derivatives, options, etc. Accuracy is a paramount concern, and even small inaccuracies can get magnified when large sums of money are involved. Even a 0.01% error in pricing can be disastrous when speculating on billions of dollars worth of securities. The particular benchmark we study here is a Monte Carlo simulation for predicting European derivative pricing [12].

In Figure 13, we plot the accuracy of Native-32, REBits-32, and Native-64 as a function of the number of Monte Carlo paths considered. Native-64 is equivalent to GMP-1000, so we do not plot GMP-1000. The key take-away point is that REBits-32 closely tracks Native-64 (less than 0.9% error) whereas Native-32 has errors as large as 9%. As with the

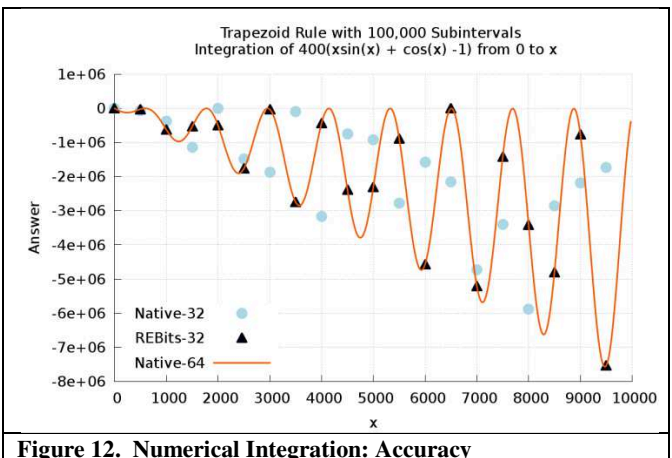

**Figure 12. Numerical Integration: Accuracy**

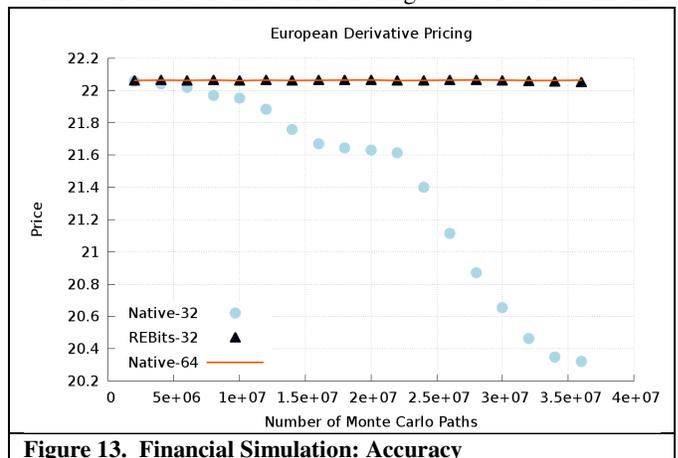

**Figure 13. Financial Simulation: Accuracy**



| Figure 14 | Figure 15 |
|---|---|
| ```
struct dd{
   double lo, hi;
}
dd_add(dd X, dd Y) {
  double s1, s2, t1, t2;
  s1 = sum(X.hi, Y.hi, &s2);
  t1 = sum(X.lo, Y.lo, &t2);t
  s2 += t1;
  s1 = quick_sum(s1, s2, &s2);
  s2 += t2;
  s1 = quick_sum(s1, s2, &s2);
  return dd(s1, s2);
}
double quick_sum(double a, double b, double &err) {
  double s = a + b;
  *err = b - (s - a); // infer error
  return s;
}
double sum(double a, double b, double &err){
  double s = a + b;
  double bb = s - a;
  *err = (a-(s-bb)) + (b-bb); // infer error
  return s;
}
``` | ```
dd_add(dd X, dd Y){
   double s1, s2, t1, t2;
   s1 = X.hi + Y.hi
   s2 = FPERR64
   t1 = X.lo + Y.lo
   t2 = FPERR64
   s2  += t1
   s1 = s1 + s2
   s2 = FPERR64
   s2 += t2
   s1 = s1 + s2
   s2 = FPERR64
   return dd(s1, s2)
}
``` |

**Figure 14. Native Double-Double addition** | **Figure 15. Double-Double addition with REBits**

numerical integration benchmark, on a processor without a 64-bit FPU, REBits-32 is an attractive option. Looking at the runtime and energy results, provided in Table 7, we find that all three schemes have almost identical costs, with differences that are within the "noise." This benchmark, like the numerical integration benchmark, has a dataset that fits in the cache and thus the differences between schemes are negligible.

## 9. REBits vs. Software Numerical Methods

### 9.1. Comparison to Software-Only Schemes

There exists a large body of prior work that uses software to improve the accuracy (not precision) of summation in the presence of finite-precision hardware. Prior work—including notable contributions by Dekker [13], Knuth [14], Kahan [15], Bailey [5], Priest [16], and Shewchuck [17]—uses additional arithmetic operations to recover the rounding errors. Retrieving the exact error takes more operations, thus incurring more overhead, which is why some schemes choose to retrieve an approximate error at lower cost. REBits can retrieve the exact error automatically in hardware.

These software schemes also tend to be non-adaptive, treating all operations as potentially worst-case (i.e., with severe cancellations and loss of significant bits). Compared to these software techniques, REBits uses a small amount of hardware to retrieve the exact error at a cost that is much lower than even the software schemes that retrieve only the approximate error. Because REBits accelerates an operation, rounding error retrieval, that is integral to many of these schemes, REBits can actually be used to accelerate some of these schemes. Table 8 summarizes the possible benefits of using REBits to accelerate these schemes. In addition, we perform a detailed case study for Double-Double in Section 9.2.

We looked at several BLAS (Basic Linear Algebra Subroutines) implementations, including Intel's MKL, because BLAS is commonly used in scientific workloads and contains many running sums. However, we found that most BLAS routines are fine-tuned for performance and offer few accuracy benefits over accuracy-unaware software. REBits can be used to improve the accuracy of BLAS without suffering significant performance degradation.

One other software-only approach, with a goal similar to REBits, would be to use 32-bit floating point numbers in memory but 64-bit numbers in the core. The programmer would declare most values to be 32-bit floats except for running sums that would be declared as 64-bit doubles. This approach is viable, but REBits is preferable in two ways. First, REBits-64 enables greater precision than that available to software. For example, REBits-64 allows the programmer to obtain nearly 128-bit floating point accuracy without a 128-bit FPU. Second, REBits-32 offers the performance of 32-bit floating point arithmetic for running sums at an accuracy close to 64-bit floating point. As mentioned earlier, the performance of 32-bit floating point performance is often twice the performance of 64-bit floating point.

### 9.2. Using REBits to Streamline Double-Double Arithmetic

We have previously (Section 3.1) discussed the Double-Double scheme [5] as a software-only approach for extending precision to achieve greater accuracy. Its accuracy is excellent, but its performance and energy consumption are many times greater than Native-64 or REBits-64. We now describe how we have used the REBits idea—recycling the error in floating point addition—to streamline Double-Double math.

Double-Double is so costly because it stitches together the results of multiple 64-bit computations on multiple 64-bit values. Consider the addition of two double-doubles, $X$ and $Y$. $X$ is two doubles ($X_{low}$ and $X_{high}$), and $Y$ is similarly two doubles ($Y_{low}$ and $Y_{high}$). To add $X$ and $Y$ to produce a double-double sum, $Z$, the Double-Double library performs the operations in Figure 14.

We observe that much of the work performed in the Double-Double code could be simplified using REBits. In Figure 15, we present functionally equivalent code that is accelerated using REBits. The code with REBits enhancement is clearly shorter and simpler. One of the key insights is that the native Double-Double code expends computational effort in inferring the error that REBits explicitly provides.

We performed similar transformations of Double-Double routines for multiplication and division. The results for all three operations are included in Table 8. We observe that using REBits can vastly reduce the computational effort—latency and energy—required for Double-Double math.

To illustrate the benefit of streamlining Double-Double math, we ran a shock hydrodynamics benchmark with Double-Double math and with our streamlined Double-Double math. The benchmark—a challenge problem in the DARPA UHPC program (https://computation.llnl.gov/casc/ShockHydro) — simulates what happens to a material when it is affected by an impulse force (e.g., a pebble dropping into a bowl of water or a bullet impacting body armor). The accuracy results are, by definition, identical, because our enhancement has no impact on functionality. The runtime and energy results in Table 9 reveal that the REBits enhancement runs for 73% as long and uses 78% as much energy.

**Table 8. Improving software numerical methods using REBits**

|  | Operation | Native Inst. with REBits | Native Instructions without REBits |
|---|---|---|---|
| **Knuth** [14] | addition | 6 fpadd | 1 fpadd, 1 move FPERR32/64 |
| **Kahan** [15] | addition | 4 fpadd, | 2 fpadd, 1 move FPERR32/64 |
| **Dekker** [13] | addition | 3 fpadd | 1 fpadd, 1 move FPERR32/64 |
| **Priest** [16] | addition | 7 fpadd, 2 fpcomp | 1 fpadd, 1 move FPERR32/64 |
| **Double-Double** [5] (All operands are in Double-Double) | addition/ subtraction | 20 fpadd | 6 fpadd, 4 move FPERR64 |
| | multiplic. | 9 fpmult 15 fpadd | 9 fpmult, 13 fpadd, 1 move FPERR64 |
| | division | 3 fpdiv 16 fpmult 81 fpadd | 3 fpdiv, 16 fpmult, 40 fpadd, 13 move FPERR64 |

**Table 9. Shock Hydrodynamics: Runtime and Energy**

|  | Runtime (s) | Energy (J) |
|---|---|---|
| Double-Double | 2835 | 55873 |
| REBits Double-Double | 2067 | 43455 |

## 10. Prior Work

We are unaware of any prior work that provides lightweight, IEEE-754-compliant architectural support for extending precision, but there is work in a few related areas.

Architectural support for SIMD floating point: Most modern ISAs include support for SIMD floating point computations. The x86 ISA has SSE (and now AVX), IBM's Power ISA has AltiVec, and ARM's ISA has NEON. SIMD provides greater performance and energy-efficiency, but it has no impact on accuracy.

Co-processors for vector/matrix math: Researchers have developed co-processors for vector and matrix math [18][19][20], particularly dot-product computations. These co-processors are similar in spirit to our work, but they are far more heavyweight and complicated.

Hardware support for interval arithmetic: One method for handling rounding error is to use interval arithmetic [21], in which each nominal value is represented by an interval (often represented with two floating point numbers, the infimum and supremum). Each value's interval is kept wide enough to include any possible rounding errors. Researchers have developed hardware support for interval arithmetic (e.g., [22]), and this support is related to our work, but computing with intervals is far more complicated and energy-intensive.

## 11. Conclusions

In this paper, we have demonstrated the benefits of architectural support for energy-efficient numerical accuracy. Making the error in each floating point addition architecturally visible provides a simple-to-use "hook" for software to use to extend the precision beyond what is supported—or supported at high performance—by the hardware. Experimental results on real hardware show that REBits-32 can provide accuracy comparable to Native-64 at costs comparable to Native-32. REBits is not applicable to all floating point algorithms, but it does apply to several important classes of algorithms.

## 12. Acknowledgments

This material is based on work supported by the National Science Foundation under grant CCF-111-5367.